\documentclass[]{eptcs}
\providecommand{\event}{DCM 2010} 

\usepackage{url}
\usepackage{amsmath}
\usepackage{amssymb}
\usepackage{amsthm}
\usepackage{braket}

\newtheorem{Theorem}{Theorem}

\newcommand{\oort}{\frac{1}{\sqrt{2}}}
\newcommand{\twoVec}[2]{\begin{pmatrix}#1\\#2\end{pmatrix}}

\newcommand{\IC}{{\mathbb{C}}}

\let\oldmarginpar\marginpar
\renewcommand\marginpar[1]{\-\oldmarginpar[\raggedleft\footnotesize #1]%
{\raggedright\footnotesize #1}}

\title{Understanding the Quantum Computational Speed-up via De-quantisation}
\author{Alastair A. Abbott \quad and \quad Cristian S. Calude
\institute{Department of Computer Science\\
University of Auckland\\
Private Bag 92019, Auckland, New Zealand}\\
{\small {\tt www.cs.auckland.ac.nz/\~{}$\{$aabb009,cristian$\}$}}
}

\def\titlerunning{Understanding the Quantum Computational Speed-up via De-quantisation}
\def\authorrunning{A. A. Abbott \& C. S. Calude}

\date{\today}

\begin{document}

\maketitle

\begin{abstract}
	
	While it seems possible that quantum computers may allow for algorithms offering a computational speed-up over classical algorithms for some problems, the issue is poorly understood. We explore this computational speed-up by investigating the ability to de-quantise quantum algorithms into classical simulations of the algorithms which are as efficient in both time and space as the original quantum algorithms. 

	The process of de-quantisation helps formulate conditions to determine if a quantum algorithm provides a real speed-up over classical algorithms. These conditions can be used to  develop new quantum algorithms more effectively (by avoiding features that could allow the algorithm to be efficiently classically simulated), as well as providing the potential to create new classical algorithms (by using features which have proved valuable for quantum algorithms). 

	Results on many different methods of de-quantisations are presented, as well as a general formal definition of de-quantisation. De-quantisations employing higher-dimensional classical bits, as well as those using matrix-simulations, put emphasis on entanglement in quantum algorithms; a key result is that any algorithm in which the entanglement is bounded is de-quantisable. These methods are contrasted with the stabiliser formalism de-quantisations due to the Gottesman-Knill Theorem, as well as those which take advantage of the topology of the circuit for a quantum algorithm. 

	The benefits of the different methods are contrasted, and the importance of a range of techniques is emphasised. We further discuss some features of quantum algorithms which current de-quantisation methods do not cover.

\end{abstract}

\section{Introduction}

	 Since Feynman first introduced the concept of a quantum computer~\cite{Feynman:1982aa} and noted the apparent exponential cost to simulate general quantum systems with classical computers there has been much interest in the power of quantum computation, in particular the possibility of using quantum physics to develop algorithms which are more efficient than classical ones. Many quantum algorithms (e.g.\ Deutsch's algorithm) have been claimed to be superior to any classical one solving the same problem, only to be discovered later that this was not the case. In order to construct good quantum algorithms it is important to know what features are necessary for a quantum algorithm to be better than a classical one. Many quantum algorithms have a trivial classical counterpart: with care, all the operations in the matrix mechanical formulation of quantum mechanics can be  computed by classical means~\cite{Ekert:1998aa}.  In this paper  we review the ability to \emph{de-quantise} a quantum algorithm to obtain a classical algorithm  which is not exponentially slower in time (or larger in space) compared to the quantum algorithm, and explore when such a de-quantisation is possible.

\section{A Preliminary Example}

\subsection{The Deutsch-Jozsa Problem}\label{sec:Deutsch-Jozsa}  

The standard formulation of the Deutsch-Jozsa problem is as follows. Let $f : \{0,1\}^n \to \{0,1\}$, and suppose we are given a black-box computing $f$ with the guarantee that $f$ is either constant (i.e.\ for all $x \in \{0,1\}^n$ and some $a \in \{0,1\}: f(x) = a$) or balanced (i.e.\ $f(x) = 0$ for exactly half of the possible inputs $x \in \{0,1\}^n$). Such a function $f$ is called \emph{valid}. The Deutsch-Jozsa problem is to determine if $f$ is constant or balanced in as few black-box calls as possible. A typical classical algorithm would require $2^{n-1}$ black-box calls, while the quantum solution requires only one. 

The special case of $n=1$ was first considered by \cite{Deutsch:1985aa} and is called the Deutsch problem; this was de-quantised by Calude~\cite{Calude:2007aa}.\footnote{Apparently, in this article the term
`de-quantisation' was used for the first time.}

It is important to note that unlike Deutsch's problem, where there are exactly two balanced and two constant functions $f$, the distribution of constant and balanced functions is asymmetrical in the Deutsch-Jozsa problem. In general, there are $N=2^n$ possible input strings, each with two possible outputs ($0$ or $1$). Hence, for any given $n$ there are $2^N$ possible functions $f$.  In this finite class, exactly two functions are constant and $\binom{N}{N/2}$ are balanced. Evidently the probability that a valid function $f: \{0,1\}^n \to \{0,1\}$ is constant tends towards zero very quickly (recall that in  Deutsch-Jozsa problem, $f$ is guaranteed to be valid). Furthermore, the probability that any randomly chosen function of the $2^N$ possible functions is valid is
$(\binom{N}{N/2}+2)\cdot 2^{-N}$,
 which again tends to zero as $n$ increases. This is clearly not an ideal problem to work with, however even in this case we can gain, via de-quantisation, useful information.

\subsubsection{Quantum Solution} 

The quantum black-box we are given takes as input three qubits and is represented by the following unitary operator $U_f$, just as it was for $n=1$:
$$U_f \ket{x} \ket{y} = \ket{x} \ket{y \oplus f(x)},$$
where $x \in \{0,1\}^2$. There are sixteen possible Boolean functions. Two of these are constant, another six are balanced and the remaining eight are not valid. All these possible functions are listed in Table~\ref{table:bFuncs}.
\begin{table}[ht]
\begin{center}
\begin{tabular}{c|cc|cccccc|cccccccc}
$f(x)$ & \multicolumn{2}{c|}{Constant} & \multicolumn{6}{c|}{Balanced} & \multicolumn{8}{c}{Invalid}\\
\hline
$f(00) = $ & \phantom{.} 0 \phantom{.} & 1 & 0 & 1 & 0 & 1 & 1 & 0 & 1 & 0 & 1 & 0 & 1 & 0 & 0 & 1 \\
$f(01) = $ & 0 & 1 & 0 & 1 & 1 & 0 & 0 & 1 & 1 & 0 & 1 & 0 & 0 & 1 & 1 & 0 \\
$f(10) = $ & 0 & 1 & 1 & 0 & 1 & 0 & 1 & 0 & 1 & 0 & 0 & 1 & 1 & 0 & 1 & 0 \\
$f(11) = $ & 0 & 1 & 1 & 0 & 0 & 1 & 0 & 1 & 0 & 1 & 1 & 0 & 1 & 0 & 1 & 0 \\
\end{tabular}
\end{center}
\caption{All possible Boolean functions $f : \{0,1\}^2 \to \{0,1\}$.}
\label{table:bFuncs}
\end{table}

Evidently, half of these functions are simply the negation of another. If we let $f'(x) = f(x) \oplus 1$ and define $\ket{\pm} = \oort (\ket{0} \pm \ket{1})$, we have:
\if01
\begin{align*}
U_{f'}\ket{x}\ket{-} &= (-1)^{f'(x)}\ket{x}\ket{-}\\
	 &= -\left((-1)^{f(x)}\ket{x}\ket{-} \right)\\
	 &= - U_f \ket{x}\ket{-}.
\end{align*}
\fi
\[U_{f'}\ket{x}\ket{-} = (-1)^{f'(x)}\ket{x}\ket{-}\\
	 = -\left((-1)^{f(x)}\ket{x}\ket{-} \right)\\
	 = - U_f \ket{x}\ket{-}.\]
In this case the result obtains a global phase factor of $-1$. Since global phase factors have no physical significance to measurement (a result is obtained with probability proportional to the amplitude squared), the outputs of $U_f$ and $U_{f'}$ are physically indistinguishable.

We will present a revised form of the standard quantum solution in which we emphasise separability of the output state.
We initially prepare our system in the state $\ket{00}\ket{1}$, and then operate on it with $H^{\otimes 3}$ to get:
\begin{equation}
\label{eqn:equalSuperpos}
H^{\otimes 3}\ket{00}\ket{1} = \frac{1}{2}\sum_{x\in \{0,1\}^2}\ket{x}\ket{-} = \ket{++}\ket{-}.
\end{equation}
In the general case, after applying the $f$-cNOT gate $U_f$ we have
\begin{equation}
\label{eqn:n2fcNot}
\begin{split}
U_f \sum_{x\in \{0,1\}^2}c_x\ket{x}\ket{-} 
	& = \left[ (-1)^{f(00)}c_{00}\ket{00} + (-1)^{f(01)}c_{01}\ket{01} +  (-1)^{f(10)}c_{10}\ket{10}\right. 
	\\ & \left. \qquad \qquad \qquad \qquad \qquad \qquad \qquad \qquad\qquad + (-1)^{f(11)}c_{11}\ket{11} \right]\ket{-}.
\end{split}
\end{equation}
From the well known rule (see e.g~\cite{Jorrand:2003aa}) for the separability of 2-qubit states, we know that this state is separable if and only if $$(-1)^{f(00)}(-1)^{f(11)}c_{00}c_{11} = (-1)^{f(01)}(-1)^{f(10)}c_{01}c_{10}.$$ While there are various initial superpositions of 2-qubit states which satisfy this condition, we only need to consider the equal superposition (shown in Equation~\ref{eqn:equalSuperpos}) that is used in this algorithm. The situation is further simplified by noting that the mapping $$(-1)^{f(a)}(-1)^{f(b)} \leftrightarrow f(a) \oplus f(b)$$ is a bijection.  In this case, the separability condition reduces to $f(00) \oplus f(11) = f(01) \oplus f(10)$. By looking back at Table~\ref{table:bFuncs} it is clear this condition holds for all balanced or constant functions $f$ for $n=2$.

We can now rewrite Equation~\ref{eqn:n2fcNot} as follows:
\begin{equation}
\label{eqn:n2Uf}
U_f \ket{++} \ket{-} = \frac{\pm 1}{2} \left( \ket{0} + (-1)^{f(00) \oplus f(10)} \ket{1} \right) \left( \ket{0} + (-1)^{f(10) \oplus f(11)} \ket{1} \right) \ket{-}.
\end{equation}

\if01
Indeed, 
\begin{align*}
& \quad (-1)^{f(00)}\ket{00} + (-1)^{f(01)}\ket{01} +  (-1)^{f(10)}\ket{10} + (-1)^{f(11)}\ket{11} \\
 &=  (-1)^{f(00)}\ket{00} + (-1)^{f(00)\oplus f(10) \oplus f(11))}\ket{01} +  (-1)^{f(10)}\ket{10} + (-1)^{f(11)}\ket{11} \\
 &=  (-1)^{f(00)}\left( \ket{00} + (-1)^{f(10)\oplus f(11)}\ket{01} +  (-1)^{f(00)\oplus f(10)}\ket{10} + (-1)^{f(00)\oplus f(11)}\ket{11} \right)\\
 &= \pm \left( \ket{0} + (-1)^{f(00) \oplus f(10)} \ket{1} \right) \left( \ket{0} + (-1)^{f(10) \oplus f(11)} \ket{1} \right),
\end{align*}
as desired.
\fi

By applying a final 3-qubit Hadamard gate to project this state onto the computational basis we obtain
\begin{multline*}
$$\frac{\pm 1}{2} H^{\otimes 3} \left( \ket{0} + (-1)^{f(00) \oplus f(10)} \ket{1} \right) \left( \ket{0} + (-1)^{f(10) \oplus f(11)} \ket{1} \right) \ket{-} = \pm \ket{f(00) \oplus f(10)} \\ \otimes \ket{f(10) \oplus f(11)} \ket{1}.$$
\end{multline*}
By measuring both the first and second qubits we can determine the nature of $f$: if both qubits are measured as $0$, then $f$ is constant, otherwise $f$ is balanced. This is correct with probability-one.

\subsubsection{De-quantising the Quantum Solution} 
\label{dqqs}

Because the quantum solution contains no entanglement, the problem can be de-quantised
by embedding classical bits in complex numbers \cite{Calude:2007aa,Abbott:2009Dissertation}. The set $ \{1,i=\sqrt{-1}\}$ acts as a computational basis in the same way that $\{\ket{0},\ket{1}\}$ does for quantum computation.\footnote{While we are not labelling the basis bits `0' and `1', they represent the classical bits 0 and 1 in the same way that $\ket{0}$ and $\ket{1}$ do.}
A complex number may be written as $z=a+bi$, so  $z$ is a natural superposition of the basis in the same way that a qubit is. 

\if01
We are now given a classical black-box that computes our function $f$. 
In order to measure the output, we need a way to project our complex numbers back on to the computational basis. This is easily done by multiplying by the input so the output is either purely imaginary or purely real.

If $z=1+i$ (an equal superposition of basis states),
$$\frac{1}{2}z \times C_f(z) = 
\begin{cases}
	\frac{\pm 1}{2}z^2 =\pm i & \text{if $f$ is constant,}\\
	\frac{\pm 1}{2}z \overline{z} = \pm 1 & \text{if $f$ is balanced.}\\
\end{cases}$$ In this manner, if the output is imaginary then $f$ is constant, if it is real then $f$ is balanced. Importantly, this is a deterministic result, and in fact the sign of the output allows us to identify \emph{which} balanced or constant function $f$ is. This is something the quantum algorithm provably cannot do~\cite{Mermin:2007aa}.
\fi

We are now given a classical black-box that computes the function $f$. 
Similarly to $U_f$, the black-box  operates on two complex numbers, $C_f : \IC^2 \to \IC^2$.  Let $z_1$, $z_2 $ be complex numbers,
\begin{equation}
	\label{deQuant:eqn:2bitCf}
	C_f\twoVec{z_1}{z_2} = C_f\twoVec{a_1+b_1i}{ a_2 + b_2i} = (-1)^{f(00)} \twoVec{a_1 + (-1)^{f(00) \oplus f(10)}b_1i}{ a_2 + (-1)^{f(10) \oplus f(11)}b_2i}\raisebox{.7mm}{.}
\end{equation}
Just as in the quantum case where the output of the black-box was two qubits that can be independently measured, the output of $C_f$ is two complex numbers  that can be independently manipulated, rather than the complex number resulting from their product. 
Note, however, that in a quantum system it is impossible to measure entangled qubits independently of each other.

To simulate a Hadamard gate 
we multiply each of the complex numbers that the black-box outputs by their respective inputs.
If we let $z_1 =  z_2 = 1+i$, we obtain the following:
\begin{displaymath}
	\frac{(1+i)}{2}\times C_f \twoVec{z_1}{z_2} = \frac{(-1)^{f(00)}}{2}\times
	\begin{cases}
		\twoVec{(1+i)(1+i)}{(1+i)(1+i)} = \twoVec{i}{i} & \\ &\text{if $f$ is constant,}\\[-2ex]
		\twoVec{(1+i)(1-i)}{(1+i)(1+i)} = \twoVec{1}{i}\\ & \\[-2ex]
		\twoVec{(1+i)(1+i)}{(1+i)(1-i)} = \twoVec{i}{1} \\ & \text{if $f$ is balanced.}\\[-2ex]
		\twoVec{(1+i)(1-i)}{(1+i)(1-i)} = \twoVec{1}{1} &
	\end{cases}
\end{displaymath}

By measuring both resulting complex numbers, we can determine whether $f$ is balanced or constant with {\it certainty}. If both complex numbers are imaginary then $f$ is constant, otherwise it is balanced. In fact,  the ability to determine if the output bits are negative or positive allows us to determine the value of $f(00)$ and thus which Boolean function $f$ is.

Because the quantum solution is \emph{separable}, it is possible to write the output of the black-box as a list of two complex numbers, and hence we can find a solution equivalent to the one obtained via a quantum computation. Writing the output in this form would not have been possible if the state was not separable, and finding a classical solution in this fashion would have required a list of complex numbers exponential in the number of input qubits. Interestingly, this de-quantisation is equivalent~\cite{Abbott:2009aa} to the `physical de-quantisation' using classical photon polarisations described by Arvind in~\cite{Arvind:2001aa}.

\subsubsection{Implementing the De-quantised Solution}
\label{implement}
\if01
An alternative classical approach can be presented using two photons. If a transformation on two qubits can be written as a transformation on each qubit independently (e.g.\ $H\otimes H$) then the transformation is trivially implemented classically. It only remains to show that the 2-qubit transformation $U_f$ can be implemented classically on two photons. Equation~\ref{eqn:n2Uf} shows that the quantum black-box  $U_f$ can be written as a product of two 1-qubit gates:\footnote{So far we have been considering the case where $U_f$ operates on $n$ input qubits and one auxiliary qubit, $\ket{-}$. It has been shown (see~\cite{Collins:1998aa}) that the auxiliary qubit is not necessary if we restrict ourselves to the subspace spanned by $\ket{-}$. We have presented the algorithm with the auxiliary qubit present because it is more intuitive to think of the input-dependent phase factor being an eigenvalue of the auxiliary qubit which is kicked back. The de-quantised solutions, however, bear more resemblance to this reduced version of $U_f$ operating only on $n$ qubits.}
\begin{align*}
U_f^{(1)} \ket{+} = \frac{1}{\sqrt{2}}\left(\ket{0} + (-1)^{f(00)\oplus f(10)}\ket{1}\right),\\
U_f^{(2)} \ket{+} = \frac{1}{\sqrt{2}}\left(\ket{0} + (-1)^{f(10)\oplus f(11)}\ket{1}\right).
\end{align*}
Each of these are valid unitary operators, and the transformation describing the black-box may be written $U_f = U_f^{(1)}\otimes U_f^{(2)}$. This means that the operation of $f$ can be computed by applying a 1-qubit operation (implemented as wave-plates) to each photon independently, and thus a classical solution  is easily found. The photons need not interact with each other at any point during the algorithm, not even inside the black-box implementation.

This classical, optical method is equivalent to both the quantum solution and the previously described classical solution. The difference is in how it is represented, bringing emphasis on the fact that  the quantum solution does not take advantage of uniquely quantum behaviour and is thus classical in nature. Further, it shows that the solution can be obtained without any interaction or sharing of information between qubits.
\fi

It is only natural to ask the question: how efficiently can we physically implement the de-quantised solution
presented in Section~\ref{dqqs}?
There are different possible approaches, but  we will discuss only one.

Nuclear magnetic resonance (NMR) spectroscopy~\cite{Levitt:2008fk} exploits the spin dynamics of nuclear spin systems; it involves placing a sample in a strong external magnetic field and hence a splitting of the nuclear spin energy levels (Zeeman splitting). The corresponding resonance frequencies (Larmor frequencies) are typically of the order of hundreds of MHz. NMR spectroscopy is a rich source of information as spectra reflect interactions of nuclear spins with their electronic environment as well as interactions (couplings) between nuclear spins themselves. 

In particular, solution-state NMR has been extensively examined as a possible implementation platform for quantum computations. This approach relies on couplings between spins within molecules and the manipulation of such finite-sized spin systems with appropriate pulse sequences. For example, Shor's algorithm has been successfully implemented in a 7-qubit NMR quantum computer~\cite{Vandersypen:2001aa}.

A  new approach proposed in~\cite{Rosello:2009} uses NMR as a classical computing substrate, where interactions between spins play no role and where the dynamics of these isolated spins can be fully described by a classical vector model. The technical difficulties of instability and decoherence present in quantum computation with NMR are less of an issue in this classical approach as their major source (internuclear couplings) is absent. Three different implementations have been demonstrated to simulate logic gates and other more complicated classical circuits. By making suitable choices of input and output parameters from the parameter space describing the NMR experiment, one can achieve different types of classical computations. The available parallelism, stability and ease in implementing two-dimensional classical bits (e.g.\ based on the three-dimensional vector model, or using two different spin species) makes NMR a well-suited substrate for implementations of de-quantised solutions of quantum algorithms. Work in progress of the groups in York (UK) and Auckland (NZ) involves NMR implementations of the de-quantised algorithms for the Deutsch-Josza problem described  in Section~\ref{sec:Deutsch-Jozsa}.

\section{Benefits}
The above example allows us to enumerate a few `immediate' benefits of de-quantisation as well as some long-term possible benefits: 
\begin{itemize}
\item an example of a problem previously thought to be classically impossible to solve, was solved by `de-quantising' a quantum solution;\\[-3ex]
\item the solution is not uniform, so not ideal. It seems hard to analyse the complexity (asymptotically) of the de-quantised solution; \\[-3ex] 
\item the de-quantised solution is stronger than the original quantum one: it is deterministic and it can
distinguish between functions not only classes (balanced/constant); \\[-3ex]
\item via de-quantisation, a  new   classical computational technique was proposed;\\[-3ex]
\item the lack of entanglement\footnote{Just one type of many features which leads to de-quantisation;
} `allowed' this type of de-quantisation; \\[-3ex]
\item de-quantisation is not only theoretical: it can lead to efficient implementations.\end{itemize}

\noindent De-quantisation can be one technique (among others) used to gain a better
 understanding of complexity in quantum computation, which  can help to:

\begin{itemize} 
 	\item  understand the power and need for quantum computation;  \\[-3ex]
	\item  more clearly see where quantum speed ups potentially come from; \\[-3ex]
	\item  develop new quantum algorithms. 
	\end{itemize}

\section{De-quantisation}

Until now we have used the term `de-quantisation' in an intuitive sense, so it is time to propose a more formal definition.

In the most general sense, a quantum circuit $C_n$ for a computation operating on an $n$-qubit input can be considered a sequence of gates $G=G_{T(n)}\dots G_1$, where each gate is either a unitary gate chosen from a fixed, finite set of gates $\mathcal{G}$, or a measurement gate. We can define a quantum algorithm in a similarly general sense. A quantum algorithm $\mathcal{A}$ is an infinite, uniformly generated, sequence of quantum circuits $(C_0, C_1, \dots)$. We say the algorithm runs in time $T(n)$ if $C_n$ contains $T(n)$ gates.
	
	Many well known algorithms fall into the class BQP, which where $T(n) = \text{poly}(n)$ and $C_n = M_{T(n)}U_{T(n)-1}\dots U_1$, where the $U_i \in \mathcal{G}$ are unitary and $M_{T(n)}$ is a measurement gate~\cite{Gruska:1999aa}. In other words, measurement is the last step of the algorithm. However, the definition of a quantum computation is more general than this, and any de-quantisation should be equally able to handle intermediate measurements and any other reasonable requirements.

 A classical algorithm is a program for a probabilistic Turing machine or any other computationally equivalent
 model of classical computation. The random access program machine is a particularly useful 
 variation which operates with an infinite set of distinguishable, numbered, but unbounded registers
each of which can contain an integer. Such a program has the capability for indirect addressing (i.e.\
the contents of a register can be used as an address to specify another register), thus allowing
for optimisations based on memory indices \cite{BoolosJeffrey:2007}.
 
	A quantum algorithm $(C_0, C_1, \dots)$  running in time $T(n)$, with output probability distribution $\mathcal{P}$, given by  a classical Turing machine that computes $C_{n}$ in time $\text{poly}(n)$
can be {\it de-quantised} if there is a probabilistic universal Turing machine $U$ such that for every
computable real $\gamma >0$ there (effectively) exists  a probability distribution $\mathcal{P'}$ with $|\mathcal{P'} - \mathcal{P}|< \gamma$ such that  $U$ sampling from $\mathcal{P'}$ runs in time $\text{poly}(T(n), \log (1/\gamma))$.

\section{De-quantisation Techniques}

\subsection{Entanglement Based Methods}

One of the simplest approaches of de-quantisation arises from simulating the matrix-mechanical formalism of the state evolution. While the quantum mechanical state vector for $n$ qubits contains, in general, $2^n$ components, under certain conditions it is possible to find compact representations for the state vector which are polynomial in $n$, and this can lead to de-quantisations. 

The simplest such case is, as in the Deutsch-Jozsa example of Section~\ref{sec:Deutsch-Jozsa}, when the state vector remains separable throughout the computation. In these situations, the mathematics of the quantum algorithm can be directly simulated in an efficient manner, because both the state vector and any transformations scale polynomially in the number of qubits $n$, and thus also in classical resources. This type of de-quantisation is simple to understand and implement classically, as mentioned for the Deutsch-Jozsa problem, but is too restrictive since most quantum algorithms make use of entanglement.

However, the conditions requiring separability can be loosened. Jozsa and Linden~\cite{Jozsa:2003aa} and Vidal~\cite{Vidal:2003kx} studied the situation where entanglement is bounded throughout the computation, and the primary result is Theorem~\ref{thm:JozsaLinden}. It was also noted~\cite{Vidal:2003kx} that these results are applicable to the simulation of continuous time quantum dynamics in some many-body systems.

\begin{Theorem}[Jozsa and Linden, \cite{Jozsa:2003aa}, Vidal \cite{Vidal:2003kx}]
	\label{thm:JozsaLinden}
	Suppose $\mathcal{A}$ is a polynomial time quantum algorithm with the property that at each step in the computation on an input of $n$-qubits, no more than $p_n$ qubits are entangled. If $p_n$ is $O(\log{n})$, i.e.\ the entanglement grows no faster than logarithmically in the input size, then the quantum computation is de-quantisable.
\end{Theorem}
This is an important result for de-quantisations, but it is not directly applicable to algorithms such as those which solve the Deutsch-Jozsa or Simon's~\cite{Simon:1997aa} problems, where the algorithm must make use of a black-box. Since this `quantum oracle' is not usually in $\mathcal{G}$ or efficiently decomposable into gates from $\mathcal{G}$, we further require that the entanglement of the quantum state is bounded both before and after the application of the black-box~\cite{Abbott:2009aa}---this allows the equivalent classical black-box to be represented in an efficient form, preserving the ability to de-quantise.
\begin{Theorem}[Abbott, \cite{Abbott:2009aa}]
	\label{thm:Abbott}
	Suppose $\mathcal{A}$ is a black-box algorithm which makes use of a black-box $U_f$. If $\mathcal{A}$  satisfies the conditions for Theorem~\ref{thm:JozsaLinden} with the gate set $\mathcal{G'}=\mathcal{G} \cup U_f$, then it is de-quantisable.
\end{Theorem}

These results require good quantum algorithms to necessarily utilise unbounded entanglement if they are to have any benefit over classical algorithms, and while this was already suspected by many, the ability to utilise these results to de-quantise known algorithms can lead to surprising classical results. Another example of such an instance is with the quantum Fourier transform (QFT). While it often creates unbounded entanglement, for certain classes of input states this is not the case and the computation remains separable~\cite{Abbott:2010aa}. It is conceivable that in various problems there may be natural constraints which enforce such conditions and allow a simple de-quantisation. 

\subsection{Circuit Topology Methods}

	The study of de-quantising the QFT has led to another class of de-quantisations which, rather than focusing on the mathematical form of the operators and states, exploits various properties of the structure of the quantum circuit for the algorithm. One of the simplest such results is that of Arahanov, Landau and Makowsky~\cite{Aharonov:2007aa}. 
	They show that a slightly modified version of the QFT circuit can be expressed in a form with logarithmic bubblewidth, a visual measure closely related to treewidth.\footnote{The bubblewidth and treewidth differ by no more than poly-logarithmic factors. See~\cite{Aharonov:2007aa} for further details.} This leads to a polynomial time classical simulation computing the QFT.
	
	In a similar fashion, both Markov and Shi~\cite{Markov:2008aa} and Jozsa~\cite{Jozsa:2006uq} have explored de-quantisation of circuits by working with tensor networks and treewidth. A tensor network for a circuit associates a tensor with every operator or end of wire in the quantum circuit, and distinct indices are used for different wire segments in the circuit. The network is simulated by contracting tensors together, and results focus around the ability to do so efficiently. While the input state must be separable in order to be simulated, this formalism has the notable advantage that it will work even if entanglement is present in the algorithm. The main result~\cite{Markov:2008aa} is Theorem~\ref{thm:MarkovShi}.
	\begin{Theorem}[Markov and Shi, \cite{Markov:2008aa}]
		\label{thm:MarkovShi}
		Quantum circuits with $T$ gates and treewidth $d$ can be simulated in time polynomial in $T$ and exponential in $d$ by the method of tensor contraction for product state inputs. Hence, polynomial size circuits with logarithmic treewidth are de-quantisable for product state inputs.
	\end{Theorem}
	Jozsa further extended~\cite{Jozsa:2006uq} the set of de-quantisable circuits to those which could be arranged so that for every qubit $i$, there are only logarithmically many 2-qubit gates applied to qubits $j$ and $k$ with $j \le i \le k$.
	
	These results, along with a few others~\cite{Yoran:2006kx,Valiant:2002aa}, provide the basis of the circuit topological de-quantisations. By dealing with circuits they are able to make use of the extensive graph theoretic literature relating to properties such as the treewidth. These results have been applied to the QFT~\cite{Yoran:2007aa}, complementing the de-quantisation using entanglement based techniques. These results have the advantage that they can simulate the circuit on arbitrary product state inputs, but unlike the bounded entanglement simulations can only sample from the probability distributions; in many cases this is reasonable, but it makes understanding the role of the QFT as a `quantum subroutine' in other algorithms more difficult~\cite{Yoran:2007aa}. 
	
	It is further worth nothing that the structural methods generally produce more complicated de-quantised algorithms. This is evident in the comparison of the different types of QFT de-quan\-tisations~\cite{Abbott:2010aa}, and is a result of being overly faithful to the quantum construction which must conform to the restrictions of avoiding measurement and locality. Another example of this is the de-quantisation result of Browne~\cite{Browne:2007aa}, who realised that Niu and Griffiths' semiclassical QFT~\cite{Griffiths:1996aa} can be easily turned into a completely classical de-quantised algorithm with no loss in efficiency. This method is different from the other structural approaches as rather than primarily focusing on the internal structure, it is more a result of the ability to measure or `sample' a qubit once all transformations involving it are completed, and using this to condition the next qubit's transformations. 

\subsection{Operator Methods}

	At the other end of the spectrum from the de-quantisation techniques which follow the evolution of the state vector, are the methods which follow the evolution of the operators acting on the state---very much as is the case in the Heisenberg representation in quantum mechanics, as opposed to the Schr\"odinger representation in which the states evolve. This approach led to the well known Gottesman-Knill Theorem~\cite{Gottesman:1999aa,Aaronson:2004aa}, which provides a de-quantisation result for algorithms using only the controlled-NOT, Hadamard and Phase gates, which are generators for the Clifford group.
	\begin{Theorem}[Gottesman-Knill,~\cite{Gottesman:1999aa}]
		\label{thm:Gottesman-Knill}
		Any quantum computation which uses only gates from the Clifford group (possibly conditioned on classical bits) and measurements on the computational basis, can be de-quantised.
	\end{Theorem}
	
	While the Clifford gates are not universal, this result is in some sense surprising because it allows de-quantisation of algorithms which contain unbounded entanglement. This result is a complement to Theorem~\ref{thm:JozsaLinden}, as it indicates that a good quantum algorithm must not permit a compact description of the state \emph{or} the operators. This counters the notion that it is entanglement which provides the quantum computational advantage. The Gottesman-Knill Theorem has further been extended by Van den Nest~\cite{Nest:2009aa} by reducing them to a simplified normal form and showing that all circuits consisting of Toffoli and diagonal gates only, followed by a basis measurement, are de-quantisable. 

	Given the advantages of the two complementary (state and operator based) de-quantisations, it is natural to ask if there is some further relation between these methods. This is an area in need of more research, and understanding the relationships between de-quantisation techniques will help understand quantum computation better. \emph{It is not unreasonable to consider next an interaction picture type de-quantisation, making the best use of compact descriptions of state and operators simultaneously}.


\section{Levels of De-quantisation}
	It is interesting to note that certain de-quantisation techniques appear to be `stronger' than others~\cite{Nest:2010aa}. Since quantum computation is inherently probabilistic, the goal of the de-quantisation is primarily to classically sample from the same probability distribution. However, the sampling techniques such as the entanglement-based techniques and the Gottesman-Knill method are somewhat artificial. In these cases, the probability distribution is calculated, and then a sample is taken by classical probabilistic methods at the end of the computation. This is in contrast to tensor-network de-quantisations, in which the de-quantised algorithm is inherently probabilistic, and the probability distribution is never calculated, only sampled. While this is sufficient for de-quantisation, the amount of work being done is somewhat different. In~\cite{Nest:2010aa} it is shown that there exist circuits for which this `weaker' sampling based form of de-quantisation is possible in polynomial time, but calculating the probability distribution is $\# P$-complete and thus at least as hard as an NP-complete problem. 
	
	This result suggests we should focus our attention on sample-based de-quantisations, but this is perhaps a little premature. Even though they may be less general, the `strong' de-quantisations have the advantage that they are trivial to compose together (unlike the `weak' methods~\cite{Yoran:2007ab}), easier to implement classically, and if the de-quantised algorithm is one where the quantum solution is correct with probability-one, such as the Deutsch-Jozsa problem, the de-quantised algorithm can be made deterministic rather than probabilistic. Examining which type of de-quantisation is possible for an algorithm gives further insight into, and distinction between the power of different quantum algorithms. On the other side of the picture, this sample-based approach to de-quantisation shows much promise to be extended, and alternative probabilistic de-quantisations are being explored~\cite{Nest:2010aa}.

\section{`Where to Next?' Is the Resounding Question}

	As we have seen, a range of de-quantisation techniques with different advantages and disadvantages have been developed. These techniques give us necessary, but not sufficient, conditions which a quantum algorithm must have in order to pose a benefit over a classical algorithm. For example, we know that a good quantum algorithm must lack both a concise description of the state and the operators. However, there may exist many other properties which allow de-quantisation, and all such properties must be absent from a good quantum algorithm~\cite{Jozsa:2003aa}. Extending these conditions to necessary conditions is the final, optimistic goal, as this would allow us to understand the relation between quantum and classical complexity classes.
	
	However, since this has proven to be extremely difficult, searching for new, different properties which allow de-quantisation is a rewarding and realistic goal. Such properties are beneficial as they deepen our understanding of the power of quantum computation, and the more insight we have to this, the more effectively we can develop quantum algorithms. 

	In order to find new de-quantisation techniques, it is worth exploring other types of quantum algorithms. Current techniques have focused around the standard algorithms which primarily consist of Fourier transforms and interference. However, alternative classes of algorithms, such as those based on quantum random walks have been studied~\cite{Aharanov:1993aa,Shenvi:2003aa}. Exploring de-quantisation in these different settings could lead to new results in this area.

Comparing the complexities of a quantum  and  a classical algorithm solving the same problem is not easy. For example, a polynomial-time classical algorithm is stronger than a polynomial-time quantum algorithm solving the same problem. 
			
For `oracle-type' problems, like the Deutsch-Jozsa problem, to compare complexities means to compare the classical and quantum black-boxes. Why? Let us recall that in the Deutsch-Jozsa problem the input is a classical black-box computing a  function $f : \{0,1\}^n \to \{0,1\}$. The quantum solution {\it embeds} the classical black-box into a (more powerful) quantum black-box, capable of computing with superposition states. Formally, we have changed the problem, as we do not operate with the given data, the classical black-box, but with a modified version of this black-box. The new black-box computes the function in a higher-dimension than the original classical one. Indeed, one could argue that we could create a classical black-box which takes as input a classical version of the `equal superposition' (which is separable), and output the suitable solution to the problem. Intuitively this is cheating, as all of the complexity has been hidden within the black-box. However, it is not clear how to take into consideration this black-box complexity, and at what point we are no longer solving the same problem.

The root of the proposed de-quantised solution lies in the fact that the embedding can be done as efficiently classically as it can be quantum-mechanically. To compare the complexities of the quantum and  de-quantised solutions we ought to compare the costs/resources necessary for performing these `embeddings'. In order to understand the cost of the embedding, it seems necessary take into consideration its  physical feasibility. Consider the following: by realising that the quantum black-box is a physical object, it must take, as input, a physical resource. If the black-box could be suitably isolated and embedded into the quantum computational system, since all physics is inherently quantum mechanical, the classical black-box could reasonably be transformed into a quantum one. It is not clear to see how the same can be done to embed the black-box in a de-quantised solution. For example, the embedding in the NMR implementation is somewhat artificial as we are able to `create' the classical black-box. However, mathematically the quantum and de-quantised algorithms are identical and this apparent difference cannot be readily evaluated. So an important question is: how de we take into account the physical cost of the embedding in order to truly evaluate the complexity of the classical and de-quantised solutions?

\section{Conclusion}
We have reviewed the ability to \emph{de-quantise} a quantum algorithm to obtain a classical algorithm  which is not exponentially slower in time  compared to the quantum algorithm. The main ideas involved in de-quantisation have been  illustrated  with
the Deutsch-Jozsa problem: from re-visiting the quantum solution to the construction of the de-quantised algorithm, the identification of the `ingredient' allowing de-quantisation, a  physical implementation of the de-quantised algorithm, to benefits and open questions. A formal definition of de-quantisation was proposed and the main techniques for de-quantisation have been briefly reviewed. Finally, the discussion of open problems has ended with the main unsolved problem related to the de-quantisation of the Deutsch-Jozsa problem: how to compare the classical  and quantum black-boxes.

\section*{Acknowledgement}

We thank Matthias Bechmann, Sonny Datt and Angelika Sebald for comments that improved this paper.
This work was in part supported by UoA Summer 2010 Fellowship (Abbott) and  UoA FRDF Grant 2010  (Calude). 
\bibliographystyle{eptcs}
\bibliography{alastairBib}

\end{document}